\documentclass[aps,twocolumn,superscriptaddress,showpacs]{revtex4}%
\usepackage{amsfonts}
\usepackage{amsmath}
\usepackage{amssymb}
\usepackage{graphicx}%
\providecommand{\U}[1]{\protect\rule{.1in}{.1in}}
\begin{document}
\title{All-optical transistor based on Rydberg atom-assisted opto-mechanical system}
\author{\ Yi-Mou Liu}
\affiliation{Center for Quantum Sciences and School of Physics, Northeast Normal
University, Changchun 130117, P. R. China}
\author{\ Feng Gao}
\affiliation{College of Physics, Jilin University, Changchun 130012, P. R. China}
\author{\ Jing Wang}
\affiliation{College of Physics, Jilin University, Changchun 130012, P. R. China}
\author{\ Jin-Hui Wu}
\affiliation{Center for Quantum Sciences and School of Physics, Northeast Normal
University, Changchun 130117, P. R. China}
\affiliation{Corresponding author: jhwu@nenu.edu.cn}
\date{\today }


\begin{abstract}
We study the optical response of double optomechanical cavity system assisted
by Rydberg atomic ensembles. And atomic ensembles are only coupled with one
side cavity by a single cavity mode. It has been realized that a long-range
manipulation for optical properties of hybrid system, by controlling the
Rydberg atomic ensembles decoupled with the optomechanical cavity. Switching
on the coupling between atoms and cavity mode, the original time reversal
symmetry of double cavity structure has been broken. Based on the controlled
optical non-reciprocity, we put forward the theoretical schemes of all-optical
controlled diode, rectifier and transistor.

\end{abstract}
\maketitle

\section{Introduction}

The invention of the electronic transistor has laid the cornerstone of the
information age. With the development of quantum information technology, the
research and fabrication of optical transistor have become an important
branch. In general, the fabrication of optical transistors is based on optical
non-reciprocity\cite{[1],[2]}. Early, optical non-reciprocity has been
achieved in optical waveguides \cite{[3],[4],[5],[6]} or optical nonlinear
system \cite{[7],[8],[9]} by breaking the time-reversal symmetry. Recently,
several alternative schemes based on different principles are proposed, such
as photoacoustic effects \cite{[10],[11]}, indirect interband photon
transitions \cite{[12],[13],[14]}, space symmetric fracture structures
\cite{[15],[16],[17],[18]}, moving systems \cite{[19],[10]} and parity
time-symmetric structures \cite{[21],[22],[23]}. In addition, for the
potential applications in photonic quantum information processing, the
abilities to be integrated on a chip, non-local control at long-range
\cite{[24],[25],[26]} and operated on a single-photon level
\cite{[27],[28],[29]} are desirable features for the realization of
nonreciprocal all-photonic devices in the future. With the rapid development
of new micro integrable devices, optomechanical systems have shown enormous
potential for application in quantum information processing
\cite{[30],[31],[32]}. It has previously been shown that optomechanical
systems can be used to induce nonreciprocal effects for light
\cite{[33],[34],[35]}. Multi-mode optomechanical systems have drawn much
attention recently. Numbers of novel and interesting phenomena are noted, such
as high-fidelity quantum state transfer \cite{[36],[37],[38]}, enhanced
quantum nonlinearities \cite{[39],[40]}, phonon lasing \cite{[41],[42]},
coherent perfect absorption (CPA) \cite{[43],[44]} and so on. Especially, the
CPA could view as an inverse process of laser \cite{[45],[46]}, which provides
a new mechanism in optomechanical system for controlling optical non
reciprocity. Currently, most of investigations on atom-assisted optomechanical
cavities coupled with independent cold atoms driven by quantum cavity modes
and classical coherent control fields \cite{[47],[48],[49],[50]}. However,
Rydberg atoms, coupled by dipole-dipole interactions (DDI), have recently been
shown to be efficient nonlinear media in cavities in order to achieve
additional control freedom of optomechanical interactions and applications
\cite{[51],[52],[53]}. An essential blockade effect based upon DDI prevents
the excitation of more than one atom into a Rydberg state within a macroscopic
volume of several micrometers in radius \cite{[54],[55],[56],[57]}. Meanwhile,
based on dipole blockade, many promising proposals have been put forward for
manipulating quantum states of atoms and photons \cite{[58],[59],[60]},
simulating many-body quantum systems \cite{[61],[62]}, generating reliable
single photons \cite{[63],[64],[65]}, and revealing some novel behaviors in
EIT \cite{[66],[67],[68],[69],[70],[71],[72]}, etc.

In this paper, we study the optical response of a symmetric double-cavity
optomechanical system assisted by Rydberg atomic ensemble (Superatom SA) with
a movable mirror of perfect reflection and driven by two coupling fields and
two probe fields. The transition from ground state to excited state of an atom
is coupled by a cavity mode, and the transition from the excited state to
Rydberg state is coupled by a classical control field. There are vdW
interactions among the Rydberg atoms in the cavity or from outside the system.
So the optical response of the hybrid system can be manipulated by controling
the Rydberg excitation of SA.

We focus especially on the transmission and reflection properties of two side
weak probe fields, by switching on and off the external control. In case (I),
we find a controlled optical diode effect, that photons only pass through the
system from one direction. In case (II), we get a controlled photon rectifier,
which can change the propagation direction of the photons, instead. In case
(III) and (IV), we achieve the function of amplification in two different
ways. The first one is based on coherent perfect synthesis (CPS), and the
other is owing to FWM effect of atom-assisted opto-mechanical system. We
expect that the all-optical controlled transistor (the optical correspondence
of classical electrical transistor), which sets controller, rectification,
amplification and other functions as a whole, could be explored to build new
tunable photonic devices on quantum information networks.

\begin{figure}[th]
\centering\includegraphics[width=8.0
cm]{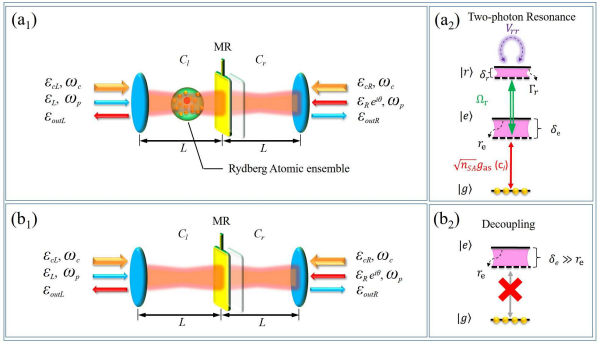}\caption{Schematic diagram of the hybrid system double
opto-mechanical cavity coupled with Rydberg atomic ensembles. (a$_{1}$) and
(a$_{2}$) show a Rydberg superatom both couple with the left cavity and the
other superatom outside. (b$_{1}$) and (b$_{2}$) show that the double-cavity
decouple with atomic ensembles swiching off the external control.}%
\label{Fig1}%
\end{figure}

\section{Theoretical Model}

We consider a hybrid optomechanical system with one movable mirror (membrane
oscillator) of perfect reflection inserted between two fixed mirrors of
partial transmission, which form two machanical-coupled Fabry--Perot cavities.
And in the left cavity there are $n_{sa}$ atoms with a Rydberg state
$|r\rangle$ [see Fig. \ref{Fig1}]. We describe the two optical modes in the
left or right cavity, respectively, by annihilation (creation) operators
$c_{l}$ $(c_{l}^{\dagger})$ or $c_{r}$ $(c_{r}^{\dagger})$. And the only
mechanical mode is described by $b$ $(b^{\dagger})$. These annihilation and
creation operators $\hat{O}=\{c_{l},c_{r},b\}$ are bosons and satisfy the
commutation relation $\left[  \hat{O}_{i},\hat{O}_{j}^{\dagger}\right]
=\delta_{i,j}$ $(i,j=c_{l},c_{r},b)$. Two probe (coupling) fields are used to
drive the double-cavity system from either left or right fixed mirrors with
their amplitudes denoted by $\varepsilon_{l}=$\textbf{ }$\sqrt{2\kappa\wp
_{L}/(\hbar\omega_{p})}$ and $\varepsilon_{r}=$\textbf{ }$\sqrt{2\kappa\wp
_{R}/(\hbar\omega_{p})}$\textbf{ (}$\varepsilon_{cL}=$\textbf{ }$\sqrt
{2\kappa\wp_{cL}/(\hbar\omega_{c})}$ and \textbf{ }$\varepsilon_{cR}=$\textbf{
}$\sqrt{2\kappa\wp_{cR}/(\hbar\omega_{c})}$\textbf{)}. Here $\kappa$ is the
decay rate of both cavity modes, $\wp_{L},\wp_{R},\wp_{cL}$ and $\wp_{cR}$ are
the relevant field powers, $\omega_{p}$ ($\omega_{c}$) is the probe (coupling)
field frequency. The membrane oscillator has an eigen frequency $\omega_{m}$
and a decay rate $\gamma_{m}$ and thus exhibits a mechanical quality factor
$Q$ = $\gamma_{m}/\omega_{m}$. Two identical optical cavities of lengths $L$
and frequencies $\omega_{0}$ are got when the membrane oscillator is at its
equilibrium position in the absence of external excitation. And in the left
Fabry--Perot cavity, cavity field $c_{l}$ also drives an ensemble of $n_{sa}$
cold atoms into the three level ladder configuration together with an external
control field with frequency $\omega_{r}$. Levels $|g\rangle$, $|e\rangle$,
and $|r\rangle$ \ correspond, respectively, to $5S_{1/2}|F=1\rangle$,
$5P_{3/2}|F=2\rangle$ and $60S_{1/2}|F=1\rangle$ of $^{87}$Rb atoms.

\bigskip Then the total Hamiltonian of our hybrid system in the rotating-wave
frame can be written as
\[
H=H_{c}+H_{m}+H_{cp}+H_{mc}+H_{a}+H_{ac}+H_{vdW},
\]
where the first four terms describe the two free optical cavities, the free
movable mirror, and the optomechanical interaction, respectively, with the
following expressions:%

\begin{align}
H_{c} &  =\mathcal{\hbar}\Delta_{c}(c_{l}^{\dagger}c_{l}+c_{r}^{\dagger}%
c_{r}),\text{ \ }H_{m}=\mathcal{\hbar}\omega_{m}b^{\dag}b,\nonumber\\
H_{cp} &  =i\mathcal{\hbar}\varepsilon_{cL}(c_{l}^{\dagger}-c_{l}%
)+i\mathcal{\hbar}(\varepsilon_{l}c_{l}^{\dagger}e^{-i\delta t}-\varepsilon
_{l}^{\ast}c_{l}e^{i\delta t})\nonumber\\
&  \mathcal{+}i\mathcal{\hbar}\varepsilon_{cR}(c_{r}^{\dagger}-c_{r}%
)+i\mathcal{\hbar}(\varepsilon_{cR}c_{r}^{\dagger}e^{i\theta}e^{-i\delta
t}-\varepsilon_{cR}^{\ast}c_{r}e^{i\theta}e^{i\delta t}),\nonumber\\
H_{mc} &  =\mathcal{\hbar}g_{0}(c_{r}^{\dagger}c_{r}-c_{l}^{\dagger}%
c_{l})(b^{\dag}+b),\label{Eq1}%
\end{align}

where we define $\Delta_{c}=\omega_{0}-\omega_{m}$ the detuning between cavity
modes and coupling fields, $\delta=\omega_{p}-\omega_{c}$ the detuning between
probe fields and coulping fields, $\theta$ the relative phase between left-
and right-side probe fields, $g_{0}=\frac{\omega_{0}}{L}\sqrt{\hbar
/(2m\omega_{m})}$ the hybrid coupling constant between mechanical and optical
modes ($m$ quality of the oscillator, $V$ the volume of the cavity, and
$\varepsilon_{0}$) the vacuum dielectric constant. The free atomic ensemble
and the atom-light interaction Hamiltonian are $H_{a}=\mathcal{\hbar}%
\sum\limits_{i=1}^{n_{sa}}(\omega_{eg}\sigma_{ee}^{(i)}+\omega_{rg}\sigma
_{rr}^{(i)})$ and $H_{ac}=-\hbar\sum\limits_{i=1}^{n_{sa}}(\Omega_{r}%
\sigma_{re}^{(i)}e^{-i\omega_{r}t}+g_{ac}c_{l}\sigma_{eg}^{(i)}+H.C.)$,
respectively, with coupling constant between atomic ensemble and left cavity
mode $g_{ac}=\wp_{ge}\sqrt{\omega_{c}/(2\hbar V\varepsilon_{0})}$ and atomic
transition dipole moment $\wp_{ge}$. We will focus on the new optomechanical
features resulting from DDI as expressed by a vdW potential%

\[
H_{vdW}=\mathcal{\hbar}\sum\limits_{i<j}^{n_{sa}}\frac{C_{6}}{R_{ij}^{6}%
}\sigma_{rr}^{(i)}\sigma_{rr}^{(j)}.
\]
Here $n_{sa}$ is the atomic number in the microvolume such that only one atom
can be excited to the state $|r\rangle$, with $C_{6}$ denoting the vdW
coefficient. And $R_{ij}$ is the interatomic distance between the $i$th and
$j$th atoms in the SA.

Considering dissipation and quantum noise of the system, we can get the
following Heisenberg-Langevin equations:
\begin{widetext}%
\begin{align}
\dot{b} &  =-i\omega_{m}b-ig_{0}[c_{l}^{\dag}c_{l}-c_{r}^{\dag}c_{r}%
]-\frac{\gamma_{m}}{2}b+\sqrt{\gamma_{m}}b^{in},\nonumber\\
\dot{c}_{l} &  =-[\kappa+i\Delta_{c}-ig_{0}(b^{\dag}+b)]c_{l}+\varepsilon
_{cL}+\varepsilon_{l}e^{i\delta t}-ig_{ac}\sqrt{n_{sa}}\tilde{P}+\sqrt
{2\kappa}c_{l}^{in},\nonumber\\
\dot{c}_{r} &  =-[\kappa+i\Delta_{c}+ig_{0}(b^{\dag}+b)]c_{r}+\varepsilon
_{cR}+\varepsilon_{r}e^{i\theta}e^{-i\delta t}+\sqrt{2\kappa}c_{r}%
^{in},\nonumber\\
\dot{P} &  =-[i\Delta_{1}+\gamma_{e}]\hat{P}+i\Omega_{r}^{\ast}\hat{S}%
+ig_{ac}\tilde{c}_{l}+f_{1}(t),\nonumber\\
\dot{S} &  =-[i\Delta_{2}+i\left\langle V\right\rangle +\gamma_{r}]\hat
{S}+i\Omega_{r}\hat{P}+f_{2}(t),\label{Eq2}%
\end{align}
\end{widetext}
where $\hat{\sigma}_{ge}=\sum_{i=1}^{n_{sa}}\hat{\sigma}_{ge}^{(i)\ }=\hat
{P}e^{-i\omega_{c}t}$ ($\hat{\sigma}_{gr}=\sum_{i=1}^{n_{sa}}\hat{\sigma}%
_{gr}^{(i)\ }=\hat{S}e^{-i(\omega_{c}+\omega_{r})t}$) denotes the collective
polarization (spin) operator of atoms, with $\tilde{c}_{l}=c_{l}e^{i\omega
_{c}t}$ and $\tilde{P}=\hat{P}e^{-i\omega_{c}t}$. $\kappa,$ $\gamma_{m},$
$\gamma_{e}$ and $\gamma_{r}$ correspond to the decay rate of cavities,
oscillator, excited state $\left\vert e\right\rangle $ and that of Rydberg
state $\left\vert r\right\rangle $, respectively. $\left\langle V\right\rangle
=$ $\sum_{i=1}^{n_{sa}}v^{(i)}$ is the mean value of total vdW potential in a
superatom with $v^{(i)}$ the vdW potential of single atom. $b^{in}$ being the
thermal noise on the movable (zero mean value), and $c_{l}^{in}(c_{r}^{in})$
is the input quantum vacuum noise from the left (right) cavity with zero mean
value. Here we are more interested in the mean optical response of the
optomechanical system to probe field in the presence of both strong driving
fields. In this regard, we can safely ignore the quantum fluctuations of all
relevant operators and use the factorization assumption $b\langle c_{i}%
\rangle=\langle b\rangle\langle c_{i}\rangle$ to generate the mean values in
steady state. In order to solve Eqs. (2), we write each operator as the sum of
its mean value and its small fluctuation $\hat{O_{1}}=\langle\hat{O_{1}%
}\rangle+\delta\hat{O_{1}},\hat{O_{1}}=\{b,c_{l},c_{r},\hat{P},\hat{S}\},$
when both coupling fields are sufficiently strong. Then we get two series of
equations about steady-state or fluctuation of operators, respectively,

\begin{widetext}%
\begin{align}
\dot{b_{s}} &  =-i\omega_{m}b_{s}-ig_{0}[c_{ls}^{\ast}c_{ls}-c_{rs}^{\ast
}c_{rs}]-\frac{\gamma_{m}}{2}b_{s},\nonumber\\
\dot{c}_{ls} &  =-[\kappa+i\Delta_{c}-ig_{0}(b_{s}^{\ast}+b_{s})]c_{ls}%
+\varepsilon_{c_{L}}-ig_{ac}\sqrt{n_{sa}}\tilde{P},\nonumber\\
\dot{c}_{rs} &  =-[\kappa+i\Delta_{c}+ig_{0}(b_{s}^{\ast}+b_{s})]c_{rs}%
+\varepsilon_{c_{R}},\nonumber\\
\dot{P_{s}} &  =-[i\Delta_{1}+\gamma_{e}]P_{s}+i\Omega_{r}^{\ast}S_{s}%
+ig_{ac}\sqrt{n_{sa}}\tilde{c}_{ls},\nonumber\\
\dot{S_{s}} &  =-[i\Delta_{2}+i\left\langle V\right\rangle +\gamma_{r}%
]S_{s}+i\Omega_{r}P_{s},\label{Eq3}%
\end{align}
\end{widetext}

\begin{widetext}
\begin{align}
\delta\dot{b} &  =-i\omega_{m}\delta b-ig_{0}[c_{ls}^{\ast}\delta c_{l}%
+c_{ls}\delta c_{l}^{\dag}-c_{rs}^{\ast}\delta c_{r}-c_{rs}\delta c_{r}^{\dag
}]-\frac{\gamma_{m}}{2}\delta b+\sqrt{\gamma_{m}}b^{in},\nonumber\\
\delta\dot{c_{l}} &  =-[\kappa+i\tilde{\Delta_{c}}]\delta c_{l}-ig_{0}%
c_{ls}(\delta b^{\dag}+\delta b)+\varepsilon_{l}e^{-i\delta t}-ig_{ac}%
\sqrt{n_{sa}}\delta\tilde{P}+\sqrt{2\kappa}c_{l}^{in},\nonumber\\
\delta\dot{c_{r}} &  =-[\kappa+i\tilde{\Delta_{c}}]\delta c_{r}+ig_{0}%
c_{rs}(\delta b^{\dag}+\delta b)+\varepsilon_{R}e^{i\theta}e^{-i\delta t}%
\sqrt{2\kappa}c_{r}^{in},\nonumber\\
\delta\hat{P} &  =-[i\Delta_{1}+\gamma_{e}]\delta\hat{P}+i\Omega_{r}^{\ast
}\delta\hat{S}+ig_{ac}\sqrt{n_{sa}}\delta\tilde{c}_{l}+f_{1}(t),\nonumber\\
\delta\hat{S} &  =-[i\Delta_{2}+iV^{s}+\gamma_{r}]\delta\hat{S}+i\Omega
_{r}\delta\hat{P}+f_{2}(t),\label{Eq3}%
\end{align}
\end{widetext}
where the effective detuning is $\tilde{\Delta_{c}}=\Delta_{c}-g_{o}b_{s}$ and
$c_{is}=\langle c_{i}\rangle$ ($i=l,r$) is the steady state average of cavities.

\bigskip It is difficult to solve Eqs. (3) and (4) directly. So we first solve
the atomic part of the equations above, taking the part of the cavities as a
constant. In general, the optical response of the Rydberg atomic ensembles are
affected by the vdW interactions. And in our system, the vdW shift reads,
\begin{align}
V^{s}  &  =\left\langle \sum_{j}^{n_{sa}}\frac{C_{6}}{R_{j}^{6}}\sigma
_{rr}^{(j)}\right\rangle .\nonumber
\end{align}

Note that $V^{s}$ tends to infinite for one Rydberg excitation $\left\langle
\hat{\sigma}_{rr}\right\rangle =1$ or vanishing for zero Rydberg excitation
$\left\langle \hat{\sigma}_{rr}\right\rangle =0$. Then Eqs.~(3) cannot be
solved straightforwardly owing to the dipole-dipole correlation between
different atoms in the rigid blockade regime. Therefore we choose to define
the $n_{sa}$ cold atoms as a single SA whose volume $V_{sa}$ of radius
$R_{sa}$ is slightly smaller than the blockade volume $V_{b}$ of radius
$R_{b}=\sqrt[6]{\frac{C_{6}\left\vert \gamma_{e}+i\Delta_{1}\right\vert
}{\left\vert \Omega_{r}\right\vert ^{2}}}$ \cite{[67]}. For a very weak cavity
field, this SA can be described by three symmetric states containing at most
one Rydberg excitation $\left\vert G\right\rangle =\left\vert g\right\rangle
^{\otimes n_{sa}},\left\vert E^{(1)}\right\rangle =\frac{1}{\sqrt{n_{sa}}}%
\sum_{i=1}^{n_{sa}}\left\vert g_{1},g_{2},...,e_{i},...,g_{n_{sa}%
}\right\rangle ,$ and $\left\vert R^{(1)}\right\rangle =\frac{1}{\sqrt{n_{sa}%
}}\sum_{i=1}^{n_{sa}}\left\vert g_{1},g_{2},...,r_{i},...,g_{n_{sa}%
}\right\rangle $ [21]. According to the Superatom theory model, the Rydberg
excitation $\Sigma_{RR}$ in an SA will determine the effective SA
polarization, which is
\[
P_{s}=\langle P_{2}\rangle\Sigma_{RR}+\langle P_{3L}\rangle\lbrack
1-\Sigma_{RR}].
\]
Here $\langle P_{2}\rangle$ is the polarization of a two level atomic system
and $\langle P_{3L}\rangle$ is that of a three level Ladder type system.
\begin{align}
\left\langle P_{2}\right\rangle  &  =\frac{ig_{ac}\sqrt{n_{sa}}\tilde{c}_{ls}%
}{i\Delta_{1}+\gamma_{e}},\\
\left\langle P_{3L}\right\rangle  &  =\frac{ig_{ac}\sqrt{n_{sa}}(i\Delta
_{2}+iV+\gamma_{r})\tilde{c}_{ls}}{(i\Delta_{1}+\gamma_{e})(i\Delta
_{2}+iV+\gamma_{r})+\Omega_{r}^{\ast}\Omega_{r}},\nonumber
\end{align}
And we get
\begin{widetext}%
\[
\Sigma_{RR}=\frac{g_{ac}^{2}n_{sa}\Omega_{r}^{\ast}\Omega_{r}\tilde{c}%
_{ls}^{\ast}\tilde{c}_{ls}}{g_{ac}^{2}n_{sa}\Omega_{c}^{\ast}\Omega_{c}%
\tilde{c}_{ls}^{\ast}\tilde{c}_{ls}+(\gamma_{e}\gamma_{r}-\Delta_{1}\Delta
_{2}+\Omega_{r}^{\ast}\Omega_{r})^{2}+\gamma_{e}^{2}\Delta_{2}^{2}+\gamma
_{r}^{2}\Delta_{1}^{2}},
\]
\end{widetext}
with $\sum_{GG}+\sum_{RR}\approx1$, neglecting the excitation of the
intermediate states of SAs.

We can get the steay-state solutions,
\begin{align}
b_{s}  &  =\frac{ig_{0}(\tilde{c}^{\ast}_{ls}\tilde{c}_{ls}-c^{\ast}%
_{rs}c_{rs})}{i\Delta_{m}+\frac{\gamma_{m}}{2}},\nonumber\\
\tilde{c}_{ls}  &  =\frac{\varepsilon_{cL}-ig_{ac}\sqrt{n_{SA}}P_{s}}%
{[\kappa+i\Delta_{c}-ig_{0}(b_{s}^{\ast}+b_{s})]},\nonumber\\
c_{rs}  &  =\frac{\varepsilon_{cR}}{[\kappa+i\Delta_{c}+ig_{0}(b_{s}^{\ast
}+b_{s})]},
\end{align}

Considering the term of crossing vdW interactions $V^{c}$, if the Rydberg
excitation $\left\langle \hat{\sigma}_{r^{\prime}r^{\prime}}\right\rangle =1$,
$(i\Delta_{2}+iV^{s}+\gamma_{r})$ ( $\gg\Omega_{c}^{2}$) will tends to be
infinite, which causes $\left\langle P_{3L}\right\rangle \approx\left\langle
P_{2}\right\rangle $ Solving equations $\partial_{t}P_{s}=-(i\Delta_{1}%
+\gamma_{e})P_{s}+i\Omega_{r}^{\ast}S_{s}+ig_{ac}\sqrt{n_{SA}}\tilde{c}%
_{ls},\partial_{t}S_{s}=-(i\Delta_{2}+iV^{s}+\gamma_{r})S_{s}+i\Omega_{r}%
P_{s}$, we can get

Using the same method, from equation $\delta\dot{\hat{P}}=-(i\Delta_{1}%
+\gamma_{e})\delta\hat{P}+i\Omega_{r}^{\ast}\delta\hat{S}+ig_{ac}\sqrt{n_{sa}%
}\delta\tilde{c}_{l}+f_{1}(t)$ and $\delta\dot{\hat{S}}=-(i\Delta
_{2}+iV+\gamma_{r})\delta\hat{S}+i\Omega_{r}\delta\hat{P}+f_{2}(t),$ we can
get \ $\left\langle \delta\hat{P}\right\rangle =\left\langle \delta\hat{P}%
_{2}\right\rangle \sum_{RR}+\left\langle \delta\hat{P}_{3L}\right\rangle
(1-\sum_{RR}),$ where
\begin{align}
\left\langle \delta\hat{P}_{2}\right\rangle  &  =\frac{ig_{ac}\sqrt{n_{sa}%
}\left\langle \delta\tilde{c}_{l}\right\rangle }{i\Delta_{1}+\gamma_{e}%
},\nonumber\\
\left\langle \delta\hat{P}_{3L}\right\rangle  &  =\frac{ig_{ac}\sqrt{n_{sa}%
}(i\Delta_{2}+iV+\gamma_{r})\left\langle \delta\tilde{c}_{l}\right\rangle
}{(i\Delta_{1}+\gamma_{e})(i\Delta_{2}+iV+\gamma_{r})+\Omega_{r}^{\ast}%
\Omega_{r}}.
\end{align}

Then keeping only the linear terms of fluctuation operators and moving into an
interaction picture by introducing: $\delta\hat{O}\rightarrow\delta\hat
{O}e^{-i\Delta_{i}t},\delta\hat{O}^{in}\rightarrow\delta\hat{O}^{in}%
e^{-i\Delta_{i}t},\hat{O}=\left\{  b,c_{l},c_{r},\hat{P},\hat{S}\right\}
,\Delta_{i}=\left\{  \Delta_{1},\Delta_{2}+V,\tilde{\Delta_{c}},\omega
_{m}\right\}  ,$we obtain the linearized quantum Langevin equations:%

\begin{align}
\delta\dot{b} &  =-ig_{0}(c_{ls}^{\ast}\delta c_{l}-c_{rs}^{\ast}\delta
c_{r})-\frac{\gamma_{m}}{2}\delta b+\sqrt{\gamma_{m}}b^{in},\nonumber\\
\delta\dot{c}_{l} &  =-\kappa\delta c_{l}-ig_{0}c_{ls}\delta b+\varepsilon
_{L}e^{ixt}-ig_{ac}\sqrt{n_{sa}}\delta\hat{P}e^{-i\omega_{c}t}+\sqrt{2\kappa
}c_{l}^{in},\nonumber\\
\delta\dot{c}_{r} &  =-\kappa\delta c_{r}+ig_{0}c_{rs}\delta b+\varepsilon
_{R}e^{i\theta}e^{-ixt}\sqrt{2\kappa}c_{r}^{in},\label{Eq6}%
\end{align}
where $x=\delta-\omega_{m}.$ And expectations of fluctuation operators satisfy
the equations:%
\begin{align}
\left\langle \delta\dot{b}\right\rangle  &  =-ig_{0}(c_{ls}^{\ast}\left\langle
\delta c_{l}\right\rangle -c_{rs}^{\ast}\left\langle \delta c_{r}\right\rangle
)-\frac{\gamma_{m}}{2}\left\langle \delta b\right\rangle ,\nonumber\\
\left\langle \delta\dot{c}_{l}\right\rangle  &  =-\kappa\left\langle \delta
c_{l}\right\rangle -ig_{0}c_{ls}\left\langle \delta b\right\rangle
+\varepsilon_{L}e^{-ixt}-ig_{ac}\sqrt{n_{sa}}\left\langle \delta\hat
{P}\right\rangle e^{-i\omega_{c}t},\nonumber\\
\left\langle \delta\dot{c}_{r}\right\rangle  &  =-\kappa\left\langle \delta
c_{r}\right\rangle +ig_{0}c_{rs}\left\langle \delta b\right\rangle
+\varepsilon_{R}e^{i\theta}e^{-ixt},\label{Eq7}%
\end{align}
where the osillating terms can be removed if steady-state solutions of Eqs.~7
are assumed to be form: $\langle\delta s\rangle=\delta s_{+}e^{-ixt}+\delta
s_{-}e^{ixt}$ with $s=b,c_{l},c_{r}$. Then it is easy to attain the following
results,
\begin{align}
\delta b_{+} &  =\frac{inGg_{AC}^{\prime}\varepsilon_{r}e^{i\theta}-iG^{\ast
}\kappa^{\prime}\varepsilon_{l}}{\gamma_{m}^{\prime}\kappa^{\prime}%
g_{AC}^{\prime}+G^{2}\kappa^{\prime}+G^{2}n^{2}g_{AC}^{\prime}},\nonumber\\
\delta c_{l+} &  =\frac{Gn\varepsilon_{r}e^{i\theta t}+[G^{2}n^{2}+\gamma
_{m}^{\prime}\kappa^{\prime}]\varepsilon_{l}}{\gamma_{m}^{\prime}%
\kappa^{\prime}g_{AC}^{\prime}+G^{2}\kappa^{\prime}+G^{2}n^{2}g_{AC}^{\prime}%
},\\
\delta c_{r+} &  =\frac{[\gamma_{m}^{\prime}g_{AC}^{\prime}+G^{2}%
]\varepsilon_{r}e^{i\theta}+G^{2}n\varepsilon_{l}}{\gamma_{m}^{\prime}%
\kappa^{\prime}g_{AC}^{\prime}+G^{2}\kappa^{\prime}+G^{2}n^{2}g_{AC}^{\prime}%
},\nonumber\label{Eq8}%
\end{align}
where we set $G=g_{0}c_{ls}$ as the effective optomechanical coupling rate and
$n^{2}=|c_{rs}/c_{ls}|^{2}$ as the photon number ratio of the two cavity
modes, with $\kappa^{\prime}=ix-\kappa,\gamma_{m}^{\prime}=ix-\frac{\gamma
_{m}}{2},g_{AC}^{\prime}=ix-\kappa-\lambda,\lambda=\frac{g_{ac}^{2}n_{sa}%
}{i\Delta_{1}+\gamma_{e}}\sum_{RR}+\frac{g_{ac}^{2}n_{sa}(i\Delta
_{2}+iV+\gamma_{r})}{(i\Delta_{1}+\gamma_{e})(i\Delta_{2}+iV+\gamma
_{r})+\Omega_{r}^{\ast}\Omega_{r}}[1-\sum_{RR}].$ In deriving Eqs.\symbol{126}%
(8), we have also assumed that $c_{ls,rs}$ is real-valued without loss of generality.

It is possible to determine the output fields $\varepsilon_{outl}$ and
$\varepsilon_{outr}$ leaving from both cavities with the following
input-output relation \cite{[44]}%

\begin{align}
\varepsilon_{outl}+\varepsilon_{l}e^{-ixt} &  =2\kappa\langle\delta
c_{l}\rangle,\nonumber\\
\varepsilon_{outr}+\varepsilon_{r}e^{i\theta}e^{-ixt} &  =2\kappa\langle\delta
c_{r}\rangle,\nonumber
\end{align}
where the oscillating terms can be removed if we set $\varepsilon
_{outl}=\varepsilon_{outl+}e^{-ixt}+\varepsilon_{outl-}e^{ixt}$ and
$\varepsilon_{outr}=\varepsilon_{outr+}e^{-ixt}+\varepsilon_{outr-}e^{ixt}.$
Note that the output components $\varepsilon_{outl+}$ and $\varepsilon
_{outr+}$ have the same Stokes frequency $\omega_{p}$ as the input probe
fields $\varepsilon_{l}$ and $\varepsilon_{r}$ while the output components
$\varepsilon_{outl-}$ and $\varepsilon_{outr-}$ are generated at the
anti-Stokes frequency $2\omega_{c}-\omega_{p}$ in a nonlinear four wave mixing
process of optomechanical interaction. Then with Eq.s (8) we can obtain
\begin{align}
\varepsilon_{outl+} &  =2\kappa\delta c_{l+}-\varepsilon_{l},\nonumber\\
\varepsilon_{outr+} &  =2\kappa\delta c_{r+}-\varepsilon_{r}e^{i\theta
},\nonumber
\end{align}
oscillating at the Stokes frequency of our special interest. So it is easy to
find
\begin{widetext}%
\begin{align}
\varepsilon_{outl+} &  =\frac{2\kappa nG^{2}\varepsilon_{r}e^{i\theta}%
+[\gamma_{m}^{\prime}\kappa(2\kappa-g_{AC}^{\prime})+G^{2}(2\kappa
n^{2}-\kappa^{\prime}-n^{2}g_{AC}^{\prime})]\varepsilon_{l}}{\gamma
_{m}^{\prime}\kappa g_{AC}^{\prime}+n^{2}g_{AC}^{\prime}G^{2}+\kappa^{\prime
}G^{2}}\nonumber\\
\varepsilon_{outr+} &  =\frac{2\kappa nG^{2}\varepsilon_{l}-[(ix-3\kappa
)(\gamma_{m}^{\prime}g_{AC}^{\prime}+G^{2})+n^{2}g_{AC}^{\prime}%
G^{2}]\varepsilon_{r}e^{i\theta}}{\gamma_{m}^{\prime}\kappa g_{AC}^{\prime
}+n^{2}g_{AC}^{\prime}G^{2}+\kappa^{\prime}G^{2}}.\nonumber
\end{align}
\end{widetext}

\section{Results and discussion}

In this section, we numerically simulate the optical response of the hybrid
system, which features controlled non-reciprocity and potential use for
optical diode and transistor. The transmission coefficient $T_{l}%
=|\varepsilon_{outr+}/\varepsilon_{l}|^{2}$ ($T_{r}=|\varepsilon
_{outl+}/\varepsilon_{r}|^{2}$) and reflection coefficient $R_{l}%
=|\varepsilon_{outl+}/\varepsilon_{l}|^{2}$ ($R_{r}=|\varepsilon
_{outr+}/\varepsilon_{r}|^{2}$) are important. And they are both the functions
of the probe frequency $x/\kappa$. In the following discussion, we only
consider the two extremes of the system: (a) optical reciprocity and (b)
optical non-reciprocity. The case (a) corresponds to decoupling between atomic
ensemble and left cavity [see Fig.2(a)]. Under the condition, we get
reflection coefficient of cavities,
\begin{widetext}%
\begin{align}
R_{l} &  =|\frac{\varepsilon_{outl+}}{\varepsilon_{l}}|^{2}=|\frac{2\kappa
n^{2}G^{2}e^{i\theta}+[\gamma_{m}^{\prime}\kappa\kappa^{\prime}+2\kappa
G^{2}-(n^{2}+1)\kappa^{\prime}G^{2}]\varepsilon_{l}/\varepsilon_{r}}%
{\gamma_{m}^{\prime}\kappa\kappa^{\prime}+(n^{2}+1)\kappa^{\prime}G^{2}}%
|^{2},\nonumber\\
R_{r} &  =|\frac{\varepsilon_{outr+}}{\varepsilon_{r}}|^{2}=|\frac{2\kappa
n^{2}G^{2}+[\gamma_{m}^{\prime}\kappa\kappa^{\prime}+2\kappa G^{2}%
-(n^{2}+1)\kappa^{\prime}G^{2}]\varepsilon_{r}e^{i\theta}/\varepsilon_{l}%
}{\gamma_{m}^{\prime}\kappa\kappa^{\prime}+(n^{2}+1)\kappa^{\prime}G^{2}}%
|^{2}.
\end{align}
\end{widetext}
The system satisfies the space reversal symmetry, because $R_{l}=R_{r}$. On
the contrary, in the case (b), the coupling between the atomic ensemble and
the cavity break the space reversal symmetry of the system. Base on the two
cases above, three functions of an optical transistor can be achieved in the
hybrid system under different conditions. We focus on the achievement of these
functions, but the external control from case (a) to case (b) will be
discussed at the end of the letter.

\begin{figure}[th]
\centering\includegraphics[width=8cm]{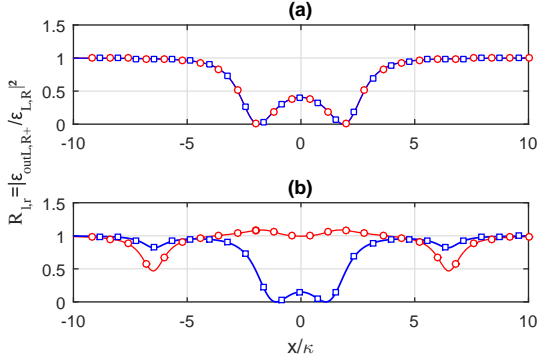}\caption{Standardization energy
of the output probe field $R_{r}=|\varepsilon_{outR+}/\varepsilon_{R}|^{2}$
(reflection coefficient of left cavity, blue) and $R_{l}=|\varepsilon
_{outL+}/\varepsilon_{L}|^{2}$ (reflection coefficient of right cavity, red
circle ) as the function of $x/\kappa$, $(a)$ switching off the external
control (coupled with atoms) and $(b)$ switching on that (decoupled with
atoms), with $G=\kappa,n=1$ and $\theta=\pi$.}%
\label{Fig2}%
\end{figure}

\subsection{Controlled Photon Diode ($\gamma=2\kappa,\theta=n\pi$ and
$\varepsilon_{l}=\varepsilon_{r}\neq0$)}

The first function achieved with the system is controlled optical diode, which
allows photons pass through only in one direction. It is coherent perfect
absorption effect (CPA) that is the basis for implementing of this function
above. The parameters should be set as $\gamma=2\kappa,\theta=n\pi$ and
$\varepsilon_{l}=\varepsilon_{r}\neq0$. CPA will occur at $x_{\pm}=\pm
\sqrt{(n^{2}+1)G^{2}-\kappa^{2}}$ \cite{[43],[44]}. However, the coupling
between atoms and the left cavity changes the CPA conditions.

In this case, Fig.~3 display the transmission coefficient of left $T_{l}$ and
right side $T_{r}$ of the hybrid system, with different $G$. It is obvious to
find the transmissions of the two sides are absolutely the same $T_{l}=T_{r}$
[see Fig.~3($a_{1}$ and $b_{1}$)], because the atomic ensemble is decoupled
with the left cavity, switching off the external control. Photons can flow
through the hybrid system in different directions, symmetrically. And then,
the optical symmetry of system will be broken ($T_{l}\neq T_{r}$),when
switching on the control causes the atoms to couple with the cavity system. In
particular, $T_{l}\simeq1$ [Fig.\symbol{126}3$(a_{2})$] in the range
$-4<x/\kappa<4$ because $g_{AC}^{\prime}>>\kappa^{\prime}$ owing to enough
large $n_{sa}^{2}$ and $|\kappa|\simeq|3\kappa-ix|<<g_{AC}^{\prime}$,
\begin{equation}
T_{l}\approx\left\vert \frac{\lbrack(3\kappa-ix)(\gamma_{m}^{\prime}%
g_{AC}^{\prime}+G^{2})+n^{2}g_{AC}^{\prime}G^{2}]}{\gamma_{m}^{\prime}\kappa
g_{AC}^{\prime}+n^{2}g_{AC}^{\prime}G^{2}+\kappa^{\prime}G^{2}}\right\vert
^{2}\approx1
\end{equation}
but $T_{r}$ also can be equal to zero [Fig.\symbol{126}3$(b_{2})$].
Furthermore, $T_{l}=1$ and $T_{r}=0$ means that we get the photon diode
framework which is an controlled unidirectional transmission device [see
Fig.\symbol{126}3($c_{1,2}$)].

\begin{figure}[th]
\centering\includegraphics[width=8cm]{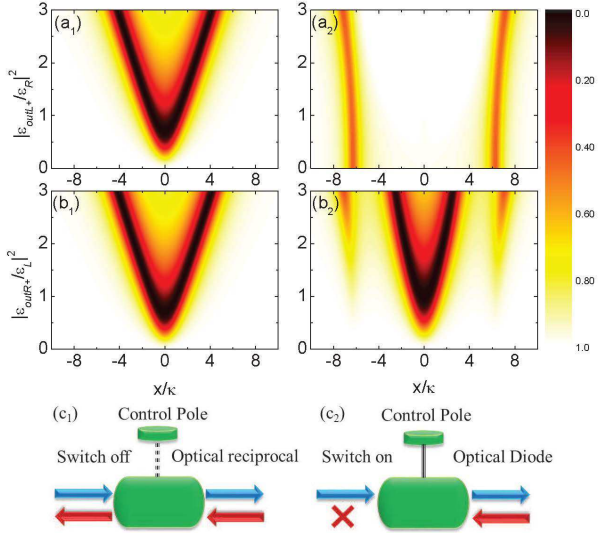}\caption{Transmission
coefficient of left side $T_{l}$ (in $a_{1}(b_{1})$) and right side $T_{r}$
(in $a_{2}(b_{2})$) of the hybrid system vs. detuning $x/\kappa$ and $G$,
without (with) external control, when $\gamma=2\kappa$ and $\theta=n\pi$ with
$n=1$. $c_{1}$ and $c_{2}$ are the functional sketch maps.}%
\label{Fig3}%
\end{figure}

\subsection{Controlled Photon Rectifier ($\gamma_{m}\longrightarrow0$,
$\theta=n\pi$, $\varepsilon_{l}\neq0$ and $\varepsilon_{r}=0$)}

Aiming to achieve optical rectification, we set $\gamma_{m}=0.1\kappa$,
$\theta=n\pi$, $\varepsilon_{l}\neq0$ and $\varepsilon_{r}=0$ (without input
field of right side). Coherent perfect transmission effect (CPT) is the basis
for implementing of this function.

In this case, if there is no coupling between atoms and cavity, we can find
$R_{l}=0$ [Fig.\symbol{126}4($a_{1}$)] and $T_{l}=1$ [Fig.\symbol{126}%
4($b_{1}$)] near the range $-0.5<x<0.5$ with $G>\kappa$. Photons flow through
the hybrid system perfectly without reflection [Fig.\symbol{126}4($c_{1}$)].
This perfect transmission is due to quantum coherence of the
double-opto-mechanical system, which can be controlled by phase $\theta$.
However, the optical properties of the system have changed, once there are
effective coupling between atoms and cavity system. We find
\begin{align}
T_{l}  &  \approx\left\vert \frac{2\kappa nG^{2}}{\gamma_{m}^{\prime}\kappa
g_{AC}^{\prime}+n^{2}g_{AC}^{\prime}G^{2}+\kappa^{\prime}G^{2}}\right\vert
^{2}\approx0,\\
R_{l}  &  \approx\left\vert \frac{\lbrack(ix-3\kappa)(\gamma_{m}^{\prime
}g_{AC}^{\prime}+G^{2})+n^{2}g_{AC}^{\prime}G^{2}]}{\gamma_{m}^{\prime}\kappa
g_{AC}^{\prime}+n^{2}g_{AC}^{\prime}G^{2}+\kappa^{\prime}G^{2}}\right\vert
^{2}\approx1,\nonumber
\end{align}
[Fig.\symbol{126}4 ($a_{2}$ and $b_{2}$)]. Photons are reflected back from the
hybrid system around the same range with $2\kappa<<g_{AC}^{\prime}$. It will
be an framework for photon rectifier, if the manipulation has been achieved to
control the direction of photon flow effectively.

\begin{figure}[th]
\centering\includegraphics[width=8cm]{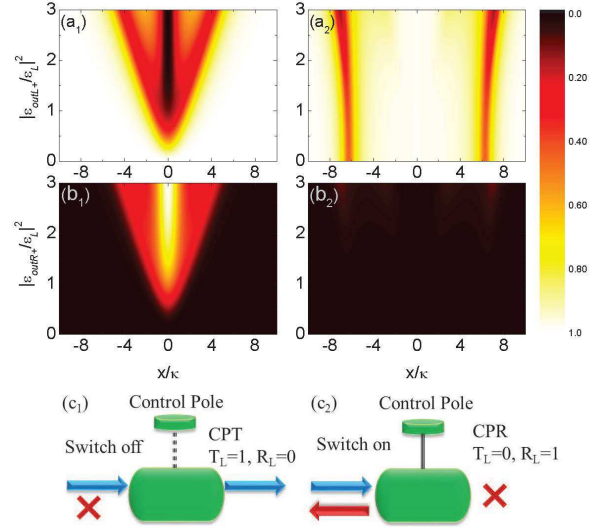}\caption{Reflection coefficient
of left side $R_{l}$ (in $a_{1}(a_{2})$) and Transmission coefficient of right
side $T_{r}$ (in $b_{1}(b_{2})$) of the hybrid system vs. detuning $x/\kappa$
and $G$, without (with) external control, when $\gamma=2\kappa$, $\gamma_{m}
\longrightarrow0$ and $\theta=n\pi$ with $\varepsilon_{l}\neq0, \varepsilon
_{r}=0$. $c_{1}$ and $c_{2}$ are the functional sketch maps only with left
side input probe field $\varepsilon_{L}$}%
\label{Fig4}%
\end{figure}

\subsection{External control}

The purpose of external control is to allow the atomic ensemble in the cavity
to be decoupled from the left cavity, controllably. We can achieve it in a
variety of ways, . Firstly, we cosider the control switching on/off the
coupling field $\Omega_{r}$, without the long range Rydberg vdW blockade
effect between two Rydberg atomic ensembles. We set $\Delta_{1}\gg g_{ac}$
with $\kappa\approx\Delta_{2}\approx0,$ which is typical Rydberg-EIT resonance
with large single photon detuning condition. Under the case, $\Omega_{r}\neq0$
is the necessary condition for effective coulping between cavity system and
Rydberg atomic ensembles, because large single photon detuning $\Delta_{1}$
will result in decoupling without $\Omega_{r}$. It is a kind of typical
coherent control, which is easy to do in the experiments. However, it is
obviously not a nonlocal or single photon level quantum control.

Secondly, we consider the control scheme that double-cavity optomechanical
system with the long range Rydberg vdW blockade effect between two Rydberg
atomic ensembles [see Fig.1(b)]. Similar to self-interactions, external vdW
interactions owing to Rydberg excitation of $\left\vert r^{\prime
}\right\rangle $ can lead to a large frequency shift of $\left\vert
r\right\rangle .$ Under the case, $i\Delta_{2}+iV+\gamma_{r}\longrightarrow
\infty$, $\left\langle P_{3L}\right\rangle \approx\left\langle P_{2}%
\right\rangle $ and we dispense with calculating $\sum_{RR}.$ With a large
single photon detuning $\Delta_{1}\gg g_{ac}$, the atoms are still decoupled
from the system. In other words, we can control the excitation of $\left\vert
r^{\prime}\right\rangle $ in the external Rydberg atomic ensembles to
manipulate the optical response of hybrid systems, which is a long range
nonlocal optical manipulation. Moreover, the control of Rydberg excitation of
the atomic ensemble outside the system can reach single-photon level.

\section{Conclusions}

\bigskip In summary, we have studied the optical response of the
Rydberg-atom-assisted double-cavity opto-mechanical system. We switch on and
off the Rydberg blockade effect to control whether the atomic ensembles
decouple with the left cavity. And whether the optical reciprocity of the
original symmetric system will be destroyed is controllable. We consider four
special cases depending on the choice of actual parameters: (I) with
$\theta=n\pi$ and $\varepsilon_{l}=\varepsilon_{r}\neq0$; (II) with
$\theta=n\pi$, $\gamma_{m}\longrightarrow0$, $\varepsilon_{l}\neq0$ and
$\varepsilon_{r}=0$.
In case (I), our numerical calculations show that atomic ensembles coupled
with single cavity break the optical reciprocity of the symmetric system.
Photons can only flow through the system in one direction, which is the
optical diode effect, switching on the control. In case (II), turning off the
input field of right side, the system has became an controlled photon
rectifier, which controls propagation behavior of the photons. In other words,
photons that should have transmitted have been reflected back absolutely.
Then, we propose two types of optical amplifier schemes with our hybrid
system. Though recent study on optical transistors have achieved great
progress, most of them have only achieved optical control. There are no the
optical correspondence of classical electric transistors which set controller,
rectification, amplification or other functions as a whole. Owing to the
blockcade effect between Rydberg atoms, it is great promising to make this
type of control to the single photon level. We hope our work can provide a new
way of thinking and a substantial role in promoting the study of optical
transistor and other all-optical devices.


\bigskip

\begin{thebibliography}{99}                                                                                               %


\bibitem {[1]}R. J. Potton, ``Reciprocity in optics,'' Rep. Prog. Phys.
\textbf{67}, 717--754 (2004).

\bibitem {[2]}I. V. Shadrivov, V. A. Fedotov, D. A. Powell, Y. S. Kivshar, and
N. I. Zheludev, ``Electromagnetic wave analogue of an electronic diode,'' New
J. Phys. \textbf{13}, 033025 (2011).

\bibitem {[3]}F. D. M. Haldane and S. Raghu, ``Possible Realization of
Directional Optical Waveguides in Photonic Crystals with Broken Time-Reversal
Symmetry,'' Phys. Rev. Lett. \textbf{100}, 013904 (2008).

\bibitem {[4]}Z. Wang, Y. Chong, J. D. Joannopoulos, and M. Soljacic,
``Observation of unidirectional backscattering-immune topological
electromagnetic states,'' Nature (London) \textbf{461}, 772--776 (2009).

\bibitem {[5]}A. B. Khanikaev, S. H. Mousavi, G. Shvets, and Y. S. Kivshar,
``One-Way Extraordinary Optical Transmission and Nonreciprocal Spoof
Plasmons,'' Phys. Rev. Lett. \textbf{105}, 126804 (2010).

\bibitem {[6]}L. Bi, J. Hu, P. Jiang, D. H. Kim, G. F. Dionne, L. C.
Kimerling, and C. A. Ross, ``On-chip optical isolation in monolithically
integrated non-reciprocal optical resonators,'' Nat. Photonics \textbf{5},
758--762 (2011).

\bibitem {[7]}K. Gallo, G. Assanto, K. R. Parameswaran, and M. M. Fejer,
``All-optical diode in a periodically poled lithium niobate waveguide,'' Appl.
Phys. Lett. \textbf{79}, 314--316 (2001).

\bibitem {[8]}L. Fan, J. Wang, L. T. Varghese, H. Shen, B. Niu, Y. Xuan, A. M.
Weiner, and M. Qi, ``An All-Silicon Passive Optical Diode,'' Science
\textbf{335}, 447--450 (2012).

\bibitem {[9]}B. Anand, R. Podila, K. Lingam, S. R. Krishnan, S. S. S. Sai, R.
Philip, and A. M. Rao, ``Optical Diode Action from Axially Asymmetric
Nonlinearity in an All-Carbon Solid-State Device,'' Nano Lett. \textbf{13},
5771--5776 (2013).

\bibitem {[10]}Q. Wang, F. Xu, Z. Y. Yu, X. S. Qian, X. K. Hu, Y. Q. Lu, and
H. T. Wang, ``A bidirectional tunable optical diode based on periodically
poled LiNbO3,'' Opt. Express \textbf{18}, 7340--7356 (2010).

\bibitem {[11]}M. S. Kang, A. Butsch, and P. S. J. Russell, ``Reconfigurable
light-driven opto-acoustic isolators in photonic crystal fibre,'' Nat.
Photonics \textbf{5}, 549--553 (2011).

\bibitem {[12]}Z. F. Yu and S. H. Fan, ``Complete optical isolation created by
indirect interband photonic transitions,'' Nat. Photonics \textbf{3}, 91--94 (2009).

\bibitem {[13]}E. Li, B. J. Eggleton, K. Fang, and S. Fan, ``Photonic
Aharonov-Bohm effect in photon-phonon interactions,'' Nat. Commun. \textbf{5},
3225 (2014).

\bibitem {[14]}M. Castellanos Munoz, A. Y. Petrov, L. OFaolain, J. Li, T. F.
Krauss, and M. Eich, ``Optically Induced Indirect Photonic Transitions in a
Slow Light Photonic Crystal Waveguide,'' Phys. Rev. Lett. \textbf{112}, 053904 (2014).

\bibitem {[15]}A. E. Miroshnichenko, E. Brasselet, and Y. S. Kivshar,
``Reversible optical nonreciprocity in periodic structures with liquid
crystals,'' Appl. Phys. Lett. \textbf{96}, 063302 (2010).

\bibitem {[16]}C. Wang, C. Zhou, and Z. Li, ``On-chip optical diode based on
silicon photonic crystal heterojunctions,'' Opt. Express \textbf{19},
26948--26955 (2011).

\bibitem {[17]}K. Xia, M. Alamri, and M. S. Zubairy, ``Ultrabroadband
nonreciprocal transverse energy flow of light in linear passive photonic
circuits,'' Opt. Express \textbf{21}, 25619--25631 (2013).

\bibitem {[18]}E. J. Lenferink, G. Wei, and N. P. Stern, ``Coherent optical
non-reciprocity in axisymmetric resonators,'' Phys. Rev. Lett. \textbf{22},
16099 (2014).

\bibitem {[19]}D. W. Wang, H. T. Zhou, M. J. Guo, J. X. Zhang, J. Evers, and
S. Y. Zhu, ``Optical Diode Made from a Moving Photonic Crystal,'' Phys. Rev.
Lett. \textbf{110}, 093901 (2013).

\bibitem {[20]}S. A. R. Horsley, J. H. Wu, M. Artoni, and G. C. La Rocca,
``Optical Nonreciprocity of Cold Atom Bragg Mirrors in Motion,'' Phys. Rev.
Lett. \textbf{110}, 223602 (2013).

\bibitem {[21]}C. Euter, K. G. Makris, R. EI-Ganainy, D. N. Christodoulides,
M. Segev, and D. Kip, ``Observation of parity-time symmetry in optics,'' Nat.
Phys. \textbf{6}, 192--195 (2010).

\bibitem {[22]}B. Peng, S. K. Ozdemir, F. Lei, F. Monifi, M. Gianfreda, G. L.
Long, S. H. Fan, F. Nori, C. M. Bender, and L. Yang, ``Parity Time-symmetric
whispering-gallery microcavities,'' Nat. Phys. \textbf{10}, 394--398 (2014).

\bibitem {[23]}J. H. Wu, M. Artoni, and G. C. La Rocca, ``Non-Hermitian
Degeneracies and Unidirectional Reflectionless Atomic Lattices,'' Phys. Rev.
Lett. \textbf{113}, 123004 (2014).

\bibitem {[24]}D. Tiarks, S. Baur, K. Schneider, S. Durr, and G. Rempe,
``Single-Photon Transistor Using a Forster Resonance,'' Phys. Rev. Lett.
\textbf{113}, 053602 (2014).

\bibitem {[25]}H. Gorniaczyk, C. Tresp, J. Schmidt, H. Fedder, and S.
Hofferberth, ``Single-Photon Transistor Mediated by Interstate Rydberg
Interactions,'' Phys. Rev. Lett. \textbf{113}, 053601 (2014).

\bibitem {[26]}W. B. Li, and I. Lesanovsky, ``Coherence in a cold-atom photon
switch,'' Phys. Rev. A \textbf{92}, 043828 (2015).

\bibitem {[27]}Y. Shen, M. Bradford, and J. T. Shen, ``Single-Photon Diode by
Exploiting the Photon Polarization in a Waveguide,'' Phys. Rev. Lett.
\textbf{107}, 173902 (2011).

\bibitem {[28]}K. Xia, G. Lu, G. Lin, Y. Cheng, Y. Niu, S. Gong, and J.
Twamley, ``Reversible nonmagnetic single-photon isolation using unbalanced
quantum coupling,'' Phys. Rev. A \textbf{90}, 043802 (2014).

\bibitem {[29]}H. Z. Shen, Y. H. Zhou, and X. X. Yi, ``Quantum optical diode
with semiconductor microcavities,'' Phys. Rev. A \textbf{90}, 023849 (2014).

\bibitem {[30]}T. J. Kippenberg and K. J. Vahala, ``Cavity Optomechanics:
Back-Action at the Mesoscale,'' Science \textbf{321}, 1172--1176 (2008).

\bibitem {[31]}F. Marquardt and S. M. Girvin, ``Trend: Optomechanics,''
Physics \textbf{2}, 40 (2009).

\bibitem {[32]}M. Aspelmeyer, T. J. Kippenberg, and F. Marquardt, ``Cavity
optomechanics,'' Rev. Mod. Phys. \textbf{86}, 1391--1452 (2014).

\bibitem {[33]}S. Manipatruni, J. T. Robinson, and M. Lipson, ``Optical
Nonreciprocity in Optomechanical Structures,'' Phys. Rev. Lett. \textbf{102},
213903 (2009).

\bibitem {[34]}M. Hafezi and P. Rabl, ``Optomechanically induced
non-reciprocity in microring resonators,'' Opt. Express \textbf{20},
7672--7684 (2012).

\bibitem {[35]}X. W. Xu and Y. Li, ``Optical nonreciprocity and optomechanical
circulator in three-mode optomechanical systems,'' Phys. Rev. A \textbf{91},
053854 (2015).

\bibitem {[36]}Y. D. Wang and A. A. Clerk, ``Using Interference for High
Fidelity Quantum State Transfer in Optomechanics,'' Phys. Rev. Lett.
\textbf{108}, 153603 (2012).

\bibitem {[37]}L. Tian, ``Adiabatic State Conversion and Pulse Transmission in
Optomechanical Systems,'' Phys. Rev. Lett. \textbf{108}, 153604 (2012).

\bibitem {[38]}H. K. Li, X. X. Ren, Y. C. Liu, and Y. F. Xiao, ``Photon-photon
interactions in a largely detuned optomechanical cavity,'' Phys. Rev. A
\textbf{88}, 053850 (2013).

\bibitem {[39]}M. Ludwig, A. H. Safavi-M. Ludwig, A. H. Safavi-Naeini, O.
Painter, and F. Marquardt, ``Enhanced Quantum Nonlinearities in a Two-Mode
Optomechanical System,'' Phys. Rev. Lett. \textbf{109}, 063601 (2012).

\bibitem {[40]}K. Stannigel, P. Komar, S. J. M. Habraken, S. D. Bennett, M. D.
Lukin, P. Zoller, and P. Rabl, ``Optomechanical Quantum Information Processing
with Photons and Phonons,'' Phys. Rev. Lett. \textbf{109}, 013603 (2012).

\bibitem {[41]}I. S. Grudinin, H. Lee, O. Painter, and K. J. Vahala, ``Phonon
Laser Action in a Tunable Two-Level System,'' Phys. Rev. Lett. \textbf{104},
083901 (2010).

\bibitem {[42]}H. Wang, Z. X. Wang, J. Zhang, S. K. Ozdemir, L. Yang, and Y.
X. Liu, ``Phonon amplification in two coupled cavities containing one
mechanical resonator,'' Phys. Rev. A \textbf{90}, 053814 (2014).

\bibitem {[43]}X. B. Yan, C. L. Cui, K. H. Gu, X. D. Tian, C. B. Fu, and J. H.
Wu, ``Coherent perfect absorption, transmission, and synthesis in a
double-cavity optomechanical system,'' Opt. Express \textbf{22}, 4886--4895 (2014).

\bibitem {[44]}G. S. Agarwal and Sumei Huang, ``Nanomechanical inverse
electromagnetically induced transparency and confinement of light in normal
modes,'' New J. Phys. \textbf{16}, 033023 (2014).

\bibitem {[45]}Y. D. Chong, Li Ge, Hui Cao, and A. D. Stone, ``Coherent
Perfect Absorbers: Time-Reversed Lasers,'' Phys. Rev. Lett. \textbf{105},
053901 (2010).

\bibitem {[46]}S. Longhi, ``Coherent perfect absorption in a homogeneously
broadened two-level medium,'' Phys. Rev. A \textbf{83}, 055804 (2011).

\bibitem {[47]}H. Ian, Z. R. Gong, Y. X. Liu, C. P. Sun, and F. Nori, ``Cavity
optomechanical coupling assisted by an atomic gas,'' Phys. Rev. A \textbf{78},
013824 (2008).

\bibitem {[48]}C. Genes, D. Vitali, and P. Tombesi, ``Emergence of
atom-light-mirror entanglement inside an optical cavity,'' Phys. Rev. A
\textbf{77}, 050307(R) (2008).

\bibitem {[49]}Y. Chang, T. Shi, Y.-X. Liu, C.-P. Sun, and F. Nori,
``Multistability of electromagnetically induced transparency in atom-assisted
optomechanical cavities,'' Phys. Rev. A \textbf{83}, 063826 (2011).

\bibitem {[50]}C. B. Fu, X. B. Yan, K. H. Gu, C. L. Cui, J. H. Wu, and T. D.
Fu, ``Steady-state solutions of a hybrid system involving atom-light and
optomechanical interactions: Beyond the weak-cavity-field approximation,''
Phys. Rev. A \textbf{87}, 053841 (2013).

\bibitem {[51]}C. Guerlin, E. Brion, T. Esslinger, and K. Molmer, ``Cavity
quantum electrodynamics with a Rydberg-blocked atomic ensemble,'' Phys. Rev. A
\textbf{82}, 053832 (2010).

\bibitem {[52]}A. Carmele, B. Vogell, K. Stannigel, and P. Zoller,
``Opto-nanomechanics strongly coupled to a Rydberg superatom: coherent versus
incoherent dynamics,'' New J. Phys. \textbf{16}, 063042 (2014).

\bibitem {[53]}A. Grankin, E. Brion, E. Bimbard, R. Boddeda, I. Usmani, A.
Ourjoumtsev, and P. Grangier, ``Quantum-optical nonlinearities induced by
Rydberg-Rydberg interactions: A perturbative approach,'' Phys. Rev. A
\textbf{92}, 043841 (2015).

\bibitem {[54]}D. Tong, S. M. Farooqi, J. Stanojevic, S. Krishnan, Y. P.
Zhang, R. Cote, E. E. Eyler, and P. L. Gould, ``Local Blockade of Rydberg
Excitation in an Ultracold Gas,'' Phys. Rev. Lett. \textbf{93}, 063001 (2004).

\bibitem {[55]}K. Singer, M. Reetz-Lamour, T. Amthor, L. G. Marcassa, and M.
Weidemuller, ``Suppression of Excitation and Spectral Broadening Induced by
Interactions in a Cold Gas of Rydberg Atoms,'' Phys. Rev. Lett. \textbf{93},
163001 (2004).

\bibitem {[56]}E. Urban, T. A. Johnson, T. Henage, L. Isenhower, D. D. Yavuz,
T. G. Walker, and M. Saffman, ``Observation of Rydberg blockade between two
atoms,'' Nat. Phys. \textbf{5}, 110--114 (2009).

\bibitem {[57]}A. Gaetan, A. Y. Miroshnychenko, T.Wilk, A. Chotia, M. Viteau,
D. Comparat, P. Pillet, A. Browaeys, and P. Grangier, ``Observation of
collective excitation of two individual atoms in the Rydberg blockade
regime,'' Nat. Phys. \textbf{5}, 115--118 (2009).

\bibitem {[58]}M. Saffman, T. G. Walker, and K. Molmer, ``Quantum information
with Rydberg atoms,'' Rev. Mod. Phys. \textbf{82}, 2313 (2010).

\bibitem {[59]}D. Jaksch, J. I. Cirac, P. Zoller, S. L. Rolston, R. Cote, and
M. D. Lukin, ``Fast Quantum Gates for Neutral Atoms,'' Phys. Rev. Lett.
\textbf{85}, 2208 (2000).

\bibitem {[60]}L. Isenhower, E. Urban, X. L. Zhang, A. T. Gill, T. Henage, T.
A. Johnson, T. G. Walker, and M. Saffman, ``Demonstration of a Neutral Atom
Controlled-NOT Quantum Gate,'' Phys. Rev. Lett. \textbf{104}, 010503 (2010).

\bibitem {[61]}H. Weimer, M. Muller, I. Lesanovsky, P. Zoller, and H. P.
Buchler, ``A Rydberg quantum simulator,'' Nat. Phys. \textbf{6}, 382--388 (2010).

\bibitem {[62]}H. Weimer, M. Muller, H. P. Buchler, and I. Lesanovsky,
``Digital quantum simulation with Rydberg atoms,'' Quantum Inf. Process.
\textbf{10}, 885--906 (2011).

\bibitem {[63]}M. Saffman and T. G. Walker, ``Creating single-atom and
single-photon sources from entangled atomic ensembles,'' Phys. Rev. A
\textbf{66}, 065403 (2002).

\bibitem {[64]}D. Porras and J. I. Cirac, ``Collective generation of quantum
states of light by entangled atoms,'' Phys. Rev. A \textbf{78}, 053816 (2008).

\bibitem {[65]}L. H. Pedersen and K. Molmer, ``Few qubit atom-light interfaces
with collective encoding,'' Phys. Rev. A \textbf{79}, 012320 (2009).

\bibitem {[66]}J. D. Pritchard, D. Maxwell, A. Gauguet, K. J. Weatherill, M.
P. A. Jones, and C. S. Adams, ``Cooperative Atom-Light Interaction in a
Blockaded Rydberg Ensemble,'' Phys. Rev. Lett. \textbf{105}, 193603 (2010).

\bibitem {[67]}D. Petrosyan, J. Otterbach, and M. Fleischhauer,
``Electromagnetically Induced Transparency with Rydberg Atoms,'' Phys. Rev.
Lett. \textbf{107}, 213601 (2011).

\bibitem {[68]}D. Yan, Y.-M. Liu, Q.-Q. Bao, C.-B. Fu, and J.-H. Wu,
``Electromagnetically induced transparency in an inverted-Ysystem of
interacting cold atoms,'' Phys. Rev. A \textbf{86}, 023828 (2012).

\bibitem {[69]}D. Yan, C.-L. Cui, Y.-M. Liu, L.-J. Song, and J.-H. Wu,
``Normal and abnormal nonlinear electromagnetically induced transparency due
to dipole blockade of Rydberg excitation,'' Phys. Rev. A \textbf{87}, 023827 (2013).

\bibitem {[70]}Y.-M. Liu, D. Yan, X.-D. Tian, C.-L. Cui, and J.-H. Wu,
``Electromagnetically induced transparency with cold Rydberg atoms: Superatom
model beyond the weak-probe approximation,'' Phys. Rev. A \textbf{89}, 033839 (2014).

\bibitem {[71]}W.-B. Li, D. Viscor, S. Hofferberth, and I. Lesanovsky,
``Electromagnetically Induced Transparency in an Entangled Medium,'' Phys.
Rev. Lett. \textbf{112}, 243601 (2014).

\bibitem {[72]}H.-Z. Wu, M.-M. Bian, L.-T. Shen, R.-X. Chen, Z.-B. Yang, and
S.-B. Zheng, ``Electromagnetically induced transparency with controlled van
der Waals interaction,'' Phys. Rev. A \textbf{90}, 045801 (2014).
\end{thebibliography}
\end{document}